\documentclass[12pt,preprint]{emulateapj}

\def\ns{RBS1774}

\def \oneight {\hbox{RX\, J1856.5-3754}}
\def \zeroseven {\hbox{RX\, J0720.4-3125}}

\def \onesix{\hbox{RX\, J1605.3+3249}}

\newcommand{\cxo}{{\it Chandra}}
\newcommand{\xmm}{{\it XMM-Newton}}
\newcommand{\vlt}{{\sl VLT}}
\newcommand{\hst}{{\sl HST}}
\newcommand{\ntt}{{\sl NTT}}
\newcommand{\vltn}{{\sl Very Large Telescope}}
\newcommand{\fors}{{\sl FORS1}}
\newcommand{\ucac}{{\sl UCAC-2}}

\newcommand{\gsc}{{\sl GSC-2}}

\slugcomment{Send offprint requests to: sz@mssl.ucl.ac.uk}

\shorttitle{An Optical Counterpart...}
\shortauthors{Zane S. et al.}

\begin{document}

\title{An Optical Counterpart Candidate for the Isolated Neutron
Star \ns}

\author{S. Zane\altaffilmark{1}, R. P. Mignani\altaffilmark{1}, R. 
Turolla\altaffilmark{2,1}, 
A. Treves\altaffilmark{3}, 
F. Haberl\altaffilmark{4}, 
C. Motch\altaffilmark{5}, 
L. Zampieri\altaffilmark{6}, 
M. Cropper\altaffilmark{1}
}
\altaffiltext{1}{
Mullard Space Science Laboratory, 
University College London
Holmbury St Mary, Dorking, Surrey, RH5 6NT,  UK
}
\altaffiltext{2}
{Department of Physics, University of Padova, via Marzolo 
8, I-35131 Padova, Italy}
\altaffiltext{3}{Universit\'a degli Studi dell'Insubria, Dipartimento di 
Fisica e
Matematica, via Valleggio 11, 22100 Como, Italy}
\altaffiltext{4}{
Max-Planck Institut f\"ur extraterrestrische Physik, 
Giessenbachstrasse, 85748 Garching, Germany}
\altaffiltext{5}
{Observatoire Astronomique,
11, rue de l'Universite, F-67000 Strasbourg, France}
\altaffiltext{6}
{INAF-Astronomical Observatory of Padova,
Vicolo dell'Osservatorio 5, I-35122 Padova, Italy}

%\offprints{sz@mssl.ucl.ac.uk}

\begin{abstract}

Multiwavelength studies of the seven identified X-ray dim isolated neutron 
stars (XDINSs) offer a unique opportunity to investigate 
their surface thermal and magnetic 
structure and the matter-radiation interaction in presence of strong 
gravitational and magnetic fields. 
As a part of an ongoing campaign aimed 
at a complete identification and spectral characterization of XDINSs in 
the 
optical band, we performed deep imaging with the ESO \vltn\ (\vlt) 
of the field of the XDINS  
\ns\ (1RXS J214303.7 +065419). The recently upgraded \fors\ instrument 
mounted on the \vlt\ provided the very first detection of a candidate 
optical counterpart in the B band. The identification is based on a very 
good positional coincidence with the X-ray source (chance probability 
$\sim 2\times 10^{-3}$). The source has B=27.4$\pm 0.2$ (1$\sigma$ 
confidence level), and the optical 
flux exceeds the extrapolation of the X-ray blackbody at optical 
wavelengths by a factor $\sim 35$ ($\pm 20$ at $3\sigma$ confidence 
level). This 
is barely compatible with thermal emission from the neutron star surface, 
unless the source distance is $d\approx 200$--300 pc, and the star is 
an almost aligned rotator or its spin axis is nearly aligned with the 
line of sight. At the same 
time, such a 
large optical excess appears difficult to reconcile with 
rotation-powered magnetospheric 
emission, unless the source has an extremely large optical emission 
efficiency. 
The implications and possible similarities with the optical 
spectra of other isolated NSs are discussed.

\end{abstract}

\keywords{star: individual (\ns) --- stars: neutron --- X-rays: stars --- 
ultraviolet: stars}

\section{Introduction}
\label{intro}

One of  the most intriguing  results of the  ROSAT All Sky  Survey has
been  the detection of  seven close-by neutron  stars (NSs), with
particular  characteristics   
\cite[XDINSs  in  the  following, see]
[for recent reviews]{ha07,vkk07}.   
These sources stand apart with  respect to  other known classes  of 
isolated NSs 
detected  at X-ray
energies.   Their X-ray spectrum is close to a blackbody, and no evidence  
of radio  emission has  been reported  so far
despite  deep searches  \cite[e.g.][]{kon08}.\footnote{The 
detection 
of pulsed 
emission
from  two   sources  has  been   claimed  at  very   low  frequencies
\citep{malo05, malo06}  but is, so  far, unconfirmed.} They  are likely to 
be endowed with 
relatively strong magnetic  fields, $B\approx  10^{13}$--$10^{14}$~G, as
inferred  from X-ray  timing  measurements and  observations of  broad
spectral lines (equivalent width $\approx 10$--100~eV, likely due to 
proton cyclotron
and/or bound-free,  bound-bound transitions  in H, H-like  and He-like
atoms). This  points  toward  a possible  evolutionary  link  between
XDINSs,    ``magnetars''   \cite[Anomalous    X-ray   Pulsars    and   
Soft Gamma-Repeaters; see][for a review]{san08}, and some of the recently 
discovered rotating radio 
transients
\cite[RRATs;][see also Heyl \& Kulkarni 1998 and Popov, Turolla \& 
Possenti~2006 for a discussion]{mcl06,mcl07}.

Detailed  multiwavelength  studies   of  XDINSs  are  fundamental  for
tracking their  evolutionary history, and for shedding  light on their
thermal and  magnetic surface  properties.  While the XDINSs  have similar
spectral properties in the X-rays, in the optical the paucity
of   multi-band   observations  prevents   a  clear   spectral
characterization. 
For the XDINSs with  a certified counterpart \cite[see e.g.][for a 
recent review]{kap08} the
optical emission lies typically a factor $\sim 10$, or more, above the
extrapolation of the X-ray  blackbody into the optical/UV band. However,
while the  optical flux closely follows  a Rayleigh-Jeans distribution
in \oneight, possible deviations from a $\lambda^{-4}$ behaviour 
have    been    reported    for    \zeroseven   \,    and    \onesix\,
\citep{kap03a,mo03,mo05,zane06}.  Thus,  whether the optical
emission from XDINSs is produced by regions  of the  star  surface at
a lower temperature \cite[e.g.][]{pons02}  or by  other mechanisms,
such as non-thermal emission  from particles in the star magnetosphere
or reprocessing  of the  surface  radiation by  an optically  thin (to 
X-rays) 
hydrogen  layer surrounding the  star \citep{mo03,za04,ho07},  is still
under debate. 
%It is therefore important to confirm and complete the 
%optical
%identifications of the remaining XDINSs.

One of the XDINSs  which so far eluded optical identification
is  \ns \,  (1RXS J214303.7 +065419).  This  source  was firstly
identified in a pointed ROSAT/PSPC observation by \cite{zamp01}; accurate 
spectral and timing information were then obtained with
\xmm\  by  \cite{za05}.   The   EPIC-PN  count   rate  (background
corrected) is $\sim 1.6$ count/s (in the 0.12-1.2 keV band), while the
0.2--2\,keV               unabsorbed              flux              is
$\sim5\times10^{-12}$~erg\,cm$^{-2}$s$^{-1}$  .  The  EPIC-PN spectrum
is very  soft and  well fitted by  an absorbed blackbody  with $kT\sim
104$  eV and  $N_H\sim 3.6\times  10^{20}$ ${\rm  cm}^{-2}$.  There is
evidence for a spectral absorption  feature at $\sim 0.7$ keV, and for
a periodicity  at 9.437 s  ($4\sigma$ confidence level) with  a pulsed
fraction  of  $\sim   4\%$  in  semiamplitude  \citep{za05,cr07}.

The first optical follow-ups with the \ntt\ and with the \vlt\ revealed no 
optical counterpart within the \xmm \ error circle, down to limiting 
magnitudes of R$\sim$22.8 \citep{zamp01,za05} and V$\sim 25.5$ 
\citep{mig07a}, respectively. Additional optical (B, V, r', i') and 
infrared (J, H and Ks) observations of \ns\ performed with the Keck, \vlt, 
Blanco and Magellan telescopes were recently reported by \cite{rea07}, 
exploiting the subarcsec position obtained through a DDT \cxo\ 
observation. Again, no plausible optical and/or infrared counterpart for 
\ns\ was detected down to r' $\sim$ 25.7 and J $\sim$ 22.6. Radio 
observations carried out with the Parkes 64m telescope at 2.9 GHz and 708 
MHz were also reported by \cite{rea07}.  However, they did not show 
evidence for radio pulsations down to a luminosity of $L =$ 0.02 mJy 
kpc$^2$ at 1.4 GHz. Very recently \cite{kon08} placed more stringent 
upper limits on the radio luminosity of RBS1774, $L_{1.4\, {\mathrm 
GHz}}\sim 0.005$ and $\sim 7.6$~mJy~kpc$^2$ for pulsed and 
bursty emission respectively, which are the most stringent limits 
obtained to date from radio observations of XDINSs.

In this  paper, we present the  first detection of  a candidate optical
counterpart to \ns, obtained with the \vlt. The   observations  and   data   
analysis   are  described   in
\S~\ref{data}, while discussion and conclusions follow in 
\S~\ref{disc} and \S~\ref{conc}, respectively.

\section{The new \vlt\ observations}
\label{data}

\subsection{Observations description}
\label{descr}

We performed deep optical imaging of the \ns\ field with \fors\ (FOcal
Reducer  Spectrograph),  a   multi-mode  instrument  for  imaging  and
long-slit/multi-object  spectroscopy  mounted   at  the  \vlt\  Kueyen
telescope  (Paranal Observatory).   The instrument  has  been recently
upgraded with  the installation of a  new detector which  is the mosaic of
two  2k$\times$4k E2V  CCDs,  optimized  for the  blue  range. Due  to
vignetting, the effective sky coverage of the two CCD chips is smaller
than the  projected detector field of  view, and larger  for the upper
chip  (dubbed ``Norma'').   Observations were  carried out  in Service
Mode on  July 11th and  21st 2007.  \fors\  was set up in  its default
standard resolution mode, with  a 2$\times$2 binning, yielding a pixel
size  of 0$\farcs$25.   The telescope  pointing  was set  in order  to
position our target  on the ``Norma'' CCD chip, and, thanks
to  its large effective  sky coverage  (7$\times$4 arcmin), to include a
large number of reference stars for a precise image astrometry. 
The low gain, fast read-out mode was chosen. 
Sequences of 590 s exposures were obtained through the B Bessel filter 
($\lambda=429 \: {\rm nm} ;  \Delta \lambda= 88 \: {\rm nm}$), for a total 
integration time of 8850 and 2950 s in the first and second night, 
respectively. The observations were collected in dark time, with an 
average seeing of $\sim 0\farcs7$ and $\sim 0\farcs9$ and an airmass $\sim 
1.17-1.37$ and $\sim 1.18$ on the first and second night, respectively .

\subsection{Data reduction}
\label{reduc}

The usual reduction  steps (bias subtraction,  flat-fielding) were applied
to    the   data    through    the   ESO    \fors\   data    reduction
pipeline\footnote{www.eso.org/observing/dfo/quality/FORS1/pipeline}
using calibration  frames acquired as  part of the  \fors\ calibration
plan.  Then, single reduced images  were aligned and coadded to filter
out cosmic rays  using the IRAF task {\tt  imcombine}. The photometric
calibration was  performed through  the observation of  standard stars
from the fields PG 1323$-$086, PG 1633+099, SA 113$-$239 \citep{lan92} at  
the  beginning  of  the night.   This  yielded  nominal
extinction  and color-corrected zero  points of  $28.16 \pm  0.04$ and
$28.17 \pm 0.04$ for the first and the second night, respectively. The
atmospheric  extinction correction  was applied  using  the extinction
coefficients  of the  Paranal Observatory  measured with  the upgraded
\fors\ ($k_B  =0.255$ for the  ``Norma'' chip).  Since the  zero point
computed by  the \fors\ pipeline is  in units of  e$^{-}$/s, while the
flux on the image is measured in ADU/s, we corrected the computed zero
point by applying the  detector electrons--to--ADU conversion factor (
{\tt GAIN=0.45}).   According to  the Paranal Observatory  sky monitor
and weather report, sky conditions were photometric on the first night
and at the  beginning of the second night but  they were then affected
by  the  presence of  thin,  variable cirri.   This means  that  the
photometry  of the  second night  is  affected by  a random,  unknown 
uncertainty. We  have tried to quantify this  uncertainty by comparing
the photometry  of a number of  reference field stars  observed in the
two  nights.   The  star   detection  was  performed  using  the  {\em
SExtractor}  program and magnitudes  were computed  through customized
aperture  photometry ({\em SExtractor}  parameter {\tt  MAG\_AUTO}). In
all sources detected at the  flux level  expected  for our  target  
(B$\ge 26$), we found an average offset of $\sim 0.1 \pm 0.5$ magnitudes 
between the photometry of the two nights. Such a large scatter for faint 
objects is due to the fact that their flux measurement is more sensitive 
on the varying atmospheric conditions than that of the bright ones. Thus, 
if data of 
the two nights are combined, this would result in only a modest 
increase of the S/N ratio at the expenses of a worse photometric accuracy 
for our target. We thus decided not to use the data taken on the second 
night.

\subsection{Astrometry}
\label{astro}

The astrometry on  the \fors\ image was computed  using as a reference
the  coordinates  of  stars  selected  from  the  \gsc\  \cite[version
3.2;][]{las08}. All  stars from  the \ucac\
\citep{zac04} are saturated in our  images.  Approximately 90 \gsc\ 
objects are
identified in the \fors\ ``Norma''  field of view. After filtering out
extended objects, stellar-like objects that are saturated or too faint
to be used as  reliable astrometric calibrators, objects falling close
to  the  chip  edges,  and  outliers,  we  performed  our  astrometric
calibration using  20 \gsc\ reference stars, evenly  distributed  in  the
instrument field of view. Their pixel coordinates
were measured  by Gaussian fitting  their intensity profiles  with the
GAIA      (Graphical      Astronomy      and      Image      Analysis)
tool\footnote{star-www.dur.ac.uk/~pdraper/gaia/gaia.html}   while  the
fit to the celestial reference  frame was performed using the Starlink
package
ASTROM\footnote{star-www.rl.ac.uk/Software/software.htm}. This code is
based on  higher  order  polynomials,  which  account  for  unmodelled  
CCD  distorsions. The
rms of the  astrometric solution was determined as $\approx$ 0\farcs11, 
per coordinate. Following Caraveo et al. (1998), we estimated the overall 
uncertainty of our astrometry  by adding in quadrature the rms of the 
astrometric fit and  the accuracy with which  we can register our field on  
the \gsc\ 
reference  frame. This is estimated as $\sqrt 3 \times \sigma_{GSC} / \sqrt N_{s}$, where  the $\sqrt 3$  term accounts  for the  free  parameters (x-scale,
y-scale, and  rotation angle) in the astrometric  fit, $\sigma_{GSC}$ is the mean  (radial) error of the \gsc\ coordinates  \cite[0\farcs3;][]{las08} and $N_{s}$ is the number of  stars used for
the astrometric  calibration. The uncertainty
on the reference stars centroids is well below 0.1 pixel and has been neglected. 
We also added in quadrature  the  radial uncertainty on the tie  of the
\gsc\  to the ICRF  \cite[0\farcs15;][]{las08}.  
Thus,  the overall  radial accuracy of  our  astrometry 
is 0\farcs24 ($1\sigma$).

\subsection{Results}
\label{res}

\begin{figure}[tb]
 \centering
 \hspace{-0.5cm}
\includegraphics[bb=40 165 555 
680,angle=-90,width=8.0cm,clip]{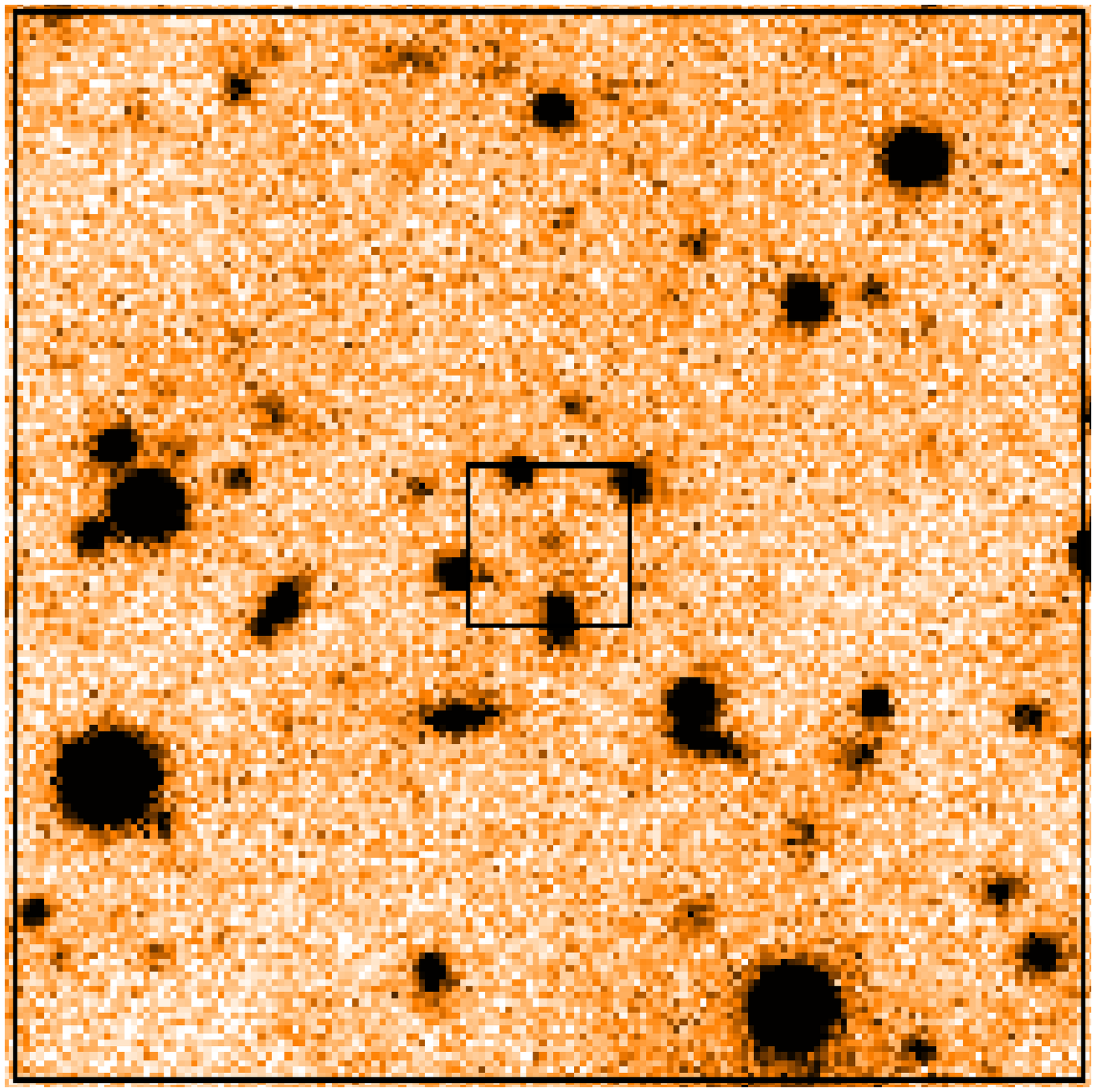}
\includegraphics[bb=50 150 550 
650,angle=-90,width=8.0cm,clip]{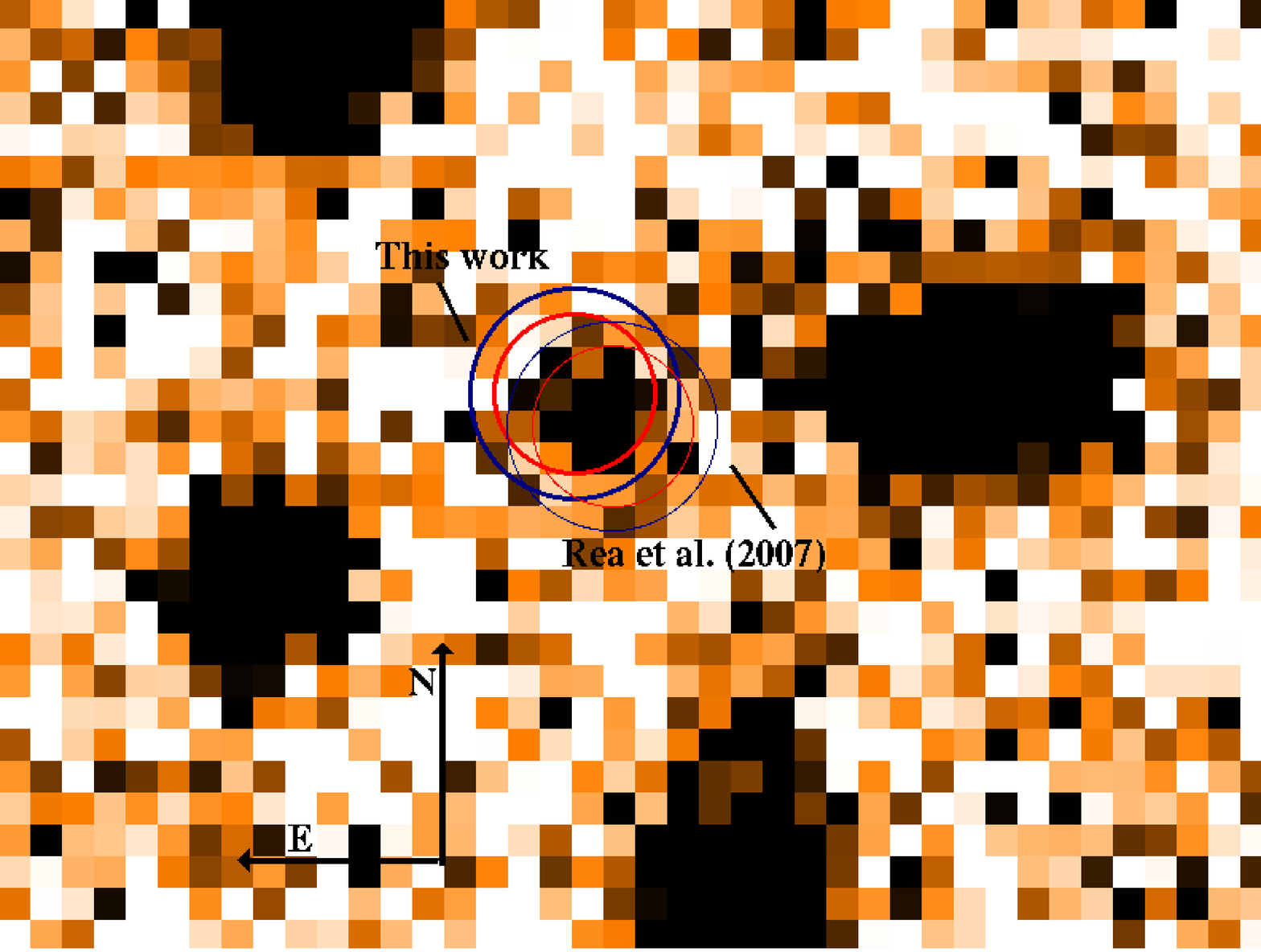}
 \caption{{\it Left.} $B$-band image ($40\arcsec \times 40\arcsec$) 
of the \ns\ field obtained with \fors\ at the \vlt\ Kueyen telescope. 
The square corresponds to the $6\arcsec \times 6 \arcsec$ zoom shown in 
the right-hand panel. The intensity scale has been re-adjusted for an 
easier view of the brighter objects in the field. {\it Right.}  
$6\arcsec \times 6 \arcsec$ zoom of the same field. 
The circles correspond to the original \citep{rea07} 
and revised (this work, \S\ref{res}) \cxo\ position of \ns, and  
are drawn at the 90\% (red) and 99\% (blue) confidence level. 
Their size (0\farcs65 and 0\farcs85, respectively) 
also accounts for the 
uncertainty of the astrometric calibration of the \fors\ image. 
The object detected at the centre of the error circles (B$=27.4 \pm 0.2$) 
is our candidate counterpart to \ns.}
\label{fig1}
\end{figure}

Fig.~\ref{fig1} shows  a section of
the co-added  \fors\ B-band image of the \ns\  field. Fig.~\ref{fig1} 
(right) shows a zoom with the \cxo\  position of \ns\  overlaied.
The coordinates of  \ns\ were measured
with high precision using  \cxo\ HRC observations  by \cite{rea07}:
$\alpha   (J2000)$=21$^h$    43$^m$   03.38$^s$,   $\delta   (J2000)$=
+06$^\circ$ 54' 17\farcs53 
with a  nominal accuracy of 0\farcs6 at the
90\%  confidence level  (c.l.).   
The \cxo\  observation was  recently
reprocessed (December  2007) using updated  calibration data products,
and revised coordinates  were obtained, $\alpha (J2000)$=21$^h$ 43$^m$
03.40$^s$,  $\delta  (J2000)$=  +06$^\circ$  54'  17\farcs79  (with 
the same  nominal accuracy; 
Israel,
private  communication),  which  are consistent  with  the
published  ones. For comparison,  in Fig.\ref{fig1}  we show  both the
original  and  the revised error circles  at  the 90\%  and  99\%
c.l.  The size  of the  error circles  also accounts  for  the overall
uncertainty  of our astrometric  calibration.  

As is evident, a  very 
faint
object  is detected about 0\farcs2 from both 
the original and revised X-ray  position. The object profile  is 
point-like and  consistent with
the  measured  PSF,  which,  at   the  level  of  the  \fors\  spatial
resolution, rules out the possibility of blending of fields object.
We computed its magnitude through aperture photometry, by using a 
customized aperture and background region, following the same procedure 
described in the previous sections. The background was computed in an 
annular region centered on the target, in order to have the most reliable 
estimate close to the position of our source. 

The  measured  value,   
corrected  for   the  atmospheric
extinction  using  the  trended  \fors\  extinction  coefficients,  is
B=27.4$\pm$ 0.2  (significant at $\sim 9 \sigma$) where the quoted  error 
($1\sigma$ c.l.) is 
purely statistical and does
not  include the  much smaller  errors on  the zero  point and  on the
atmospheric extinction correction.  No other object is detected within
or close  to the \cxo\ position down  to a $3 \sigma$  limit of B$\sim
28.7$. As pointed out in the previous sections, the \fors\ coordinates
are tied  to the ICRF  within $0\farcs15$ so that we  exclude an
hypothetical shift with respect to the \cxo\ coordinates which are also
tied to the ICRF\footnote{http://cxc.harvard.edu/cal/ASPECT/celmon/}. Thus, given its very good coincidence with the X-ray
position,  we  regard this  object as  a
viable candidate counterpart to \ns.

\section{Discussion}
\label{disc}

In this paper, we report the 
first detection of a likely optical counterpart for \ns. 
Standardly, optical identifications of isolated neutron stars are 
robustly confirmed either 
by the detection of optical pulsations or 
by the measurement of a significant proper motion. 
In absence of such information we can base  the optical 
identification only on  the positional
coincidence between  the coordinates of our  candidate counterpart and
those of  \ns, as measured  in the X-rays.   In order to  quantify the
statistical   significance,  we   estimated  the   chance  coincidence
probability that an unrelated field object might fall within the \cxo\
error circle. This  can be computed as $1  - \exp (-\pi \mu r^{2})$ (see, e.g., Severgnini et al. 2005), 
where
$\mu$ is the measured object  density in the \fors\ ``Norma'' field of
view  and $r$  is  the registered  radius  of the  \cxo\ error  circle
($0\farcs65$, accounting for the  uncertainty of the \fors\ astrometry
while  the uncertainty due to  the unknown \ns\  proper motion is
negligible  given the  small time  span  between the  \cxo\ and  \vlt\
epochs). The measured  density of stellar objects in  the field with
magnitude  27.2 $\le$B$\le$27.7, i.e.   comparable to that of our
candidate counterpart,  is $\sim$  0.0015/sq arcsec. This  yields an
estimated  chance coincidence  probability $P  \sim 2  \times 10^{-3}$
which shows  that our association  is robust.\footnote{If we compute the
chance probability by considering all objects
brighter than B=27.7, we get a very similar result, $P =
2.3\times 10^{-3}$.}
Thus, we  are confident
that we have identified a very likely optical counterpart to \ns.

\begin{figure}[tb]
 \centering
 \hspace{-0.5cm}
\includegraphics[angle=0,width=0.7\textwidth]{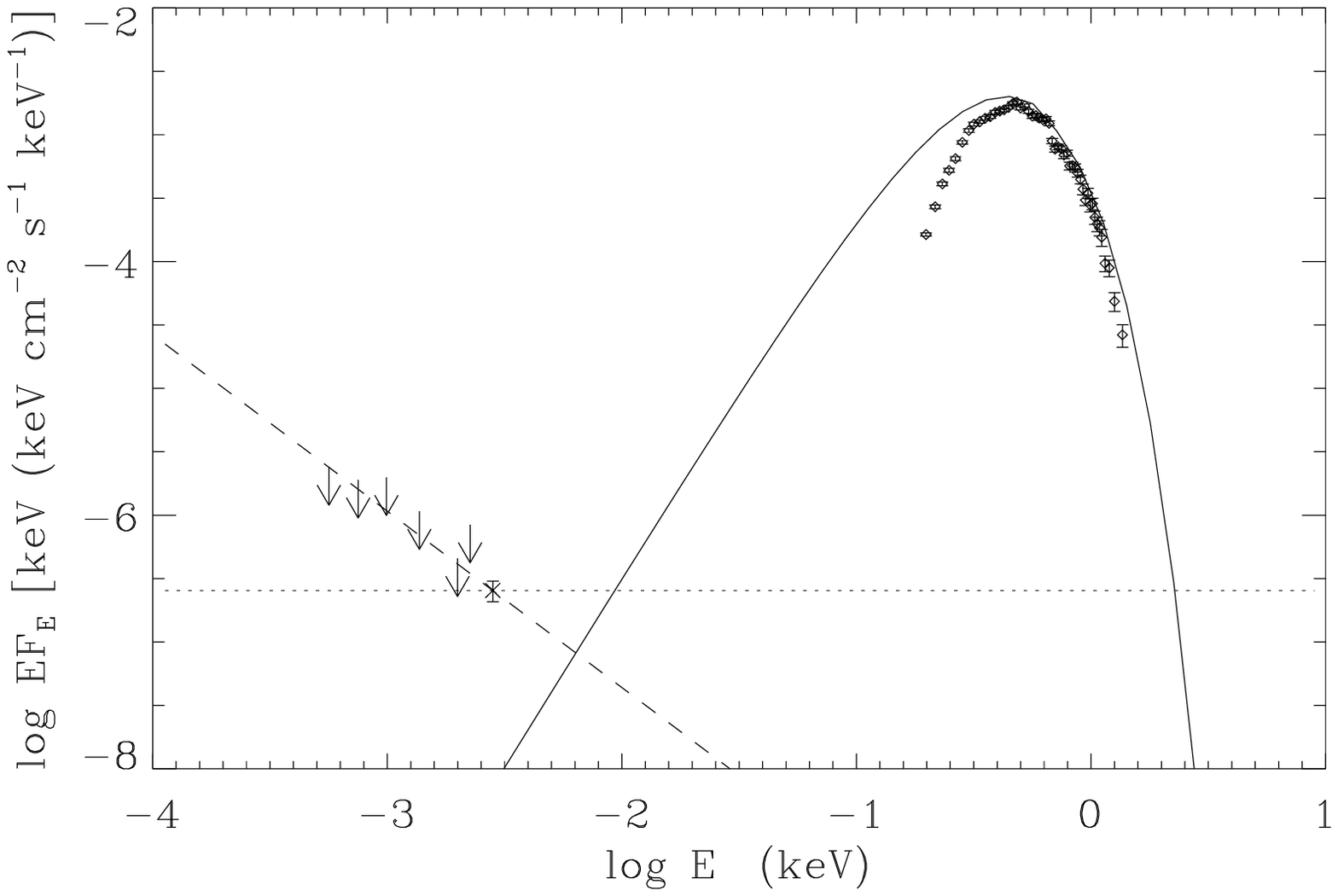}
 \caption{Multi-band spectrum of \ns. Diamonds represent
the \xmm\  spectrum \citep{za05}. The solid line shows the
unabsorbed blackbody that best fits the X-ray data, while arrows 
represent the 5$\sigma$ upper limits reported by \cite{rea07}.
The new \vlt\ measurement is shown as a cross. Dotted and dashed lines 
represent 
two powerlaw components matching the B-flux and with slope $\alpha=0, 
1.4$, respectively.  Figure re-adapted 
from \cite{rea07}.
}
 \label{fig2}
\end{figure}

The flux of our candidate counterpart in the B band, 
after correcting for interstellar extinction, is $F_{B} =
 (2.55 \pm 0.47) 
\times 10^{-7}$~keV/cm$^{2}$/s (the error is at $1\sigma$ c.l.). To 
compute this value, we used 
the column density derived from the best fit to the \xmm \ spectrum 
\cite[$N_H= 3.60 \times 10^{20}$ cm$^{-2}$,][]{cr07}, with the $A_V$
derived according to the relation of \cite{pr95}.\footnote{The \cite{pr95} 
relation is affected by 
uncertainties for close objects, due to 
the problems of modelling the ISM at small distance from the Sun where
microstructures weight more. We checked that, when using the relations of 
\cite{bo78}, as  in \cite{rea07}, and of \cite{par84} 
extinction corrections are consistent within 0.04 magnitudes, 
well below 
the
pure statistical error on the source count rate.} The interstellar 
extinction in the $B$-band has been computed
using the extinction coefficients of \cite{fitz99}. 
When compared to the extrapolation in the B band of the blackbody which 
best-fits the \xmm \  
spectrum, 
$F_{B,x} = 7.36 \times 10^{-9}$~keV/cm$^{2}$/s 
\citep{cr07}, this
gives an optical excess of $35\pm 20$ (at $3\sigma$ c.l.), where all the uncertainties on the magnitude--to--flux conversion are negligible. The
result is shown in Fig.~\ref{fig2}, where the optical/IR
upper limits of \cite{rea07} are also overplotted. The  optical 
excess 
is larger than that typically observed in other XDINSs, and this may cast 
some  doubts on the association of the newly detected source with \ns. On the 
other hand, if, as we suggest, the association is real, 
it can be used to infer physical constraints 
on the mechanism that is responsible for the optical emission.

The first scenario we consider is one in which the 
optical emission originates from a cooler fraction of the neutron
star surface, which  emits as a blackbody at temperature  $T_o$ 
\citep{br02, pons02,kap03b, tru04}\footnote{Note, however, that when 
applied to other XDINSs this scenario often requires large neutron star 
radii.}. In 
this case, the ratio
between the  optical and X-ray  fluxes scales  as $ \approx
r_o^2  T_o/r_x^2 T_x \equiv f $, where  $r_o$ is  the radial size  of the  
cold 
region 
(which of course can not exceed the  maximum value of the neutron star 
radius),
$r_x$ and $T_x$  are the blackbody radius and  temperature as inferred
from the  X-ray spectrum, $T_x=104$~eV, $r_x  = 2
(d/300 \, {\rm  pc})$~km \citep{cr07}. By making an assumption on  
$r_o$ we can obtain the lower limit on $T_o$  that corresponds to 
values of $f$ between the central value 
and the $3 \sigma$ lower limit ($15\leq f\leq 35$). 
Furthermore, since no contribution from such a cold component is 
observed in the 0.1-1~keV  \xmm\  spectrum, it must be $R \ll 1$, where 
\begin{equation}\nonumber 
R = \left (\frac{r_o}{r_x} \right)^2
 \left (\frac{T_o}{T_x} \right)^4
 \frac{ \int_{0.1/T_o}^{1/T_o} t^3/(\exp(t)-1) dt}{
 \int_{0.1/T_x}^{1/T_x} t^3/(\exp(t)-1) dt} \, . 
\nonumber 
\end{equation}
We repeated this calculation by varying the distance between $200$ and 
$500$~pc\footnote{We do not use 
the distance determination by \cite{pos07}, since their model is 
currently under revision for the case of \ns\ (Posselt, private 
communication).} and for $r_o = 20, 15$ or $10$~km. 
%;results are reported in Table~\ref{tab1}
Results show that this scenario is only barely compatible 
with our data and requires very small distances: 
$d\sim 300-400$~pc are only allowed within $3 \sigma$ 
{\it and} for neutron star radii as large as $r_o=20$~km, while 
if $r_o=15$~km the only allowed combinations 
require $f=15-22$ and $d \sim 200-250$~pc. 
On the other hand, the lower the 
distance the smaller is the size of the X-ray emitting region and the more 
difficult is to explain the small pulsed fraction 
observed in the X-rays (which is  $ \sim 4\%$ in semiamplitude). 
%This is 
%illustrated in Fig.~\ref{fig3}, where we show the PF contours 
%in the $(\chi, \xi)$ plane, where $\chi$ and $\xi$ are the inclination 
%angle of the line of sight and of the magnetic dipole axis with
%respect to the star spin axis. 
%The curves have been computed by using a simple model of blackbody 
%emission (no beaming effect is considered apart from the 
%general relativistic one) and 
%a surface thermal map consisting of two hot caps at $T_x = 
%104$~eV, centered at the magnetic poles, and with size $r_x$ as given in 
%Table~\ref{tab1}.  The rest of the surface is at a lower 
%temperature, $T_o=16$~eV. 
As we tested by assuming simple polar cap modelling and blackbody 
emission, if $d=300$~pc the allowed parameter space is already confined to 
a very small region where the inclination
angles of the line of sight and of the magnetic dipole axis with
respect to the star spin axis are both  
$\lesssim 20^\circ$. If 
$d=200$~pc, the allowed region is even smaller and the only possible 
configurations require the star to be either 
a perfectly aligned rotator or viewed almost along its spin axis. 

As  discussed by \cite{mo03},  \cite{za04} and \cite{ho07},  
spectral models
consisting of  bare neutron stars  surrounded by thin  atmospheres may
predict very different amount of optical  excess. However,  such models 
are  affected by our poor  knowledge of
the properties of the condensate surface, and whether they can produce 
optical
excesses as large as that measured here is still an open issue.

An  alternative interpretation is  that the optical  emission is
non thermal, probably of magnetospheric origin \citep{ps83}. As it can 
be seen from
Fig.~\ref{fig2}, a power law spectral component $E^{-\alpha}$ matching
the B-flux and with index  $\alpha \sim 0-1.4$ is not in contradiction
with the upper  limits measured at longer wavelengths  and, at the same
time, is  not expected to  contribute significantly to the  X-ray band.
These constraints  on the  optical 
spectral index are compatible with the values determined from
the optical  spectra of rotation-powered neutron stars,  which are all
in the range 0--0.8 \cite[see, e.g.,][]{mig07b}. 
In this respect, we note that  for the XDINSs for which the spin
down luminosity is measured \citep{cr04, kap05a, kap05b, vkk08} it is 
$\dot E \sim
5 \times 10^{30}$~erg/s.  If the case of  \ns\ is analogous, and if we  
take $L_{opt}/\dot E \approx 10^{-6}$ for 
the optical emission efficiency 
\cite[as measured in old rotation-powered neutron stars by][]{mig04}, 
this yields   
an
optical luminosity of $\sim 5 \times  10^{24}$ ~erg/s.  
In order to reproduce
the  observed flux  of   our  candidate   counterpart,  after
accounting for the assumed  interstellar extinction, \ns\ should be at
an  unrealistically small  distance of  $\sim 25$  pc.  Thus,  for the
assumed value of $\dot  E$, a purely rotation-powered optical emission
is only compatible with a scenario in which our source 
is at least a two orders of magnitude  more efficient 
optical  emitter (see below). Similar conclusions can be reached if we 
consider the case in which the observed optical emission of
\ns\ is due to 
a composite mechanism consisting  of both thermal emission from  the 
neutron star surface and a rotation-powered magnetospheric 
emission \cite[as in  the middle-aged neutron stars 
PSR~B0656+14 and Geminga, e.g.][]{kp07}. 

A further, intriguing possibility is that the optical emission is 
non-thermal and powered 
by a mechanism different from rotation. Interestingly, at least in the IR
domain,  hints have  been  found for  an  increase  of the  low-energy
emission efficiency  with the dipole magnetic  field strength 
\citep{mig07c}. 
Indeed,  in magnetars  the efficiency  is at
least two orders of  magnitude larger than in rotation-powered neutron
stars and the spectra show a typical (and unexplained) flattening  toward 
the infrared \citep{isr03}. It is  interesting to note  
the magnetic field strength of \ns\ as inferred from 
the absorption feature in the X-ray spectrum is the highest among 
all XDINSs 
($B\approx 10^{14}$~G), larger than the QED critical limit and comparable 
with that of magnetars. 
The optical emission  of \ns\ could thus be powered by
a magnetar-like process. If the spectrum of \ns\ shows a similar 
magnetar-like turn over, then the flattening is appearing blueward of  
the IR band.

\section{Conclusions}
\label{conc}

We report here the deepest optical observations so far of the \ns\ field,
performed  with the  upgraded  \fors\ instrument  at the  \vlt\ Kueyen
telescope.   Based on the positional coincidence,  we have
identified a likely  candidate counterpart.  \ns\  would then be
the fifth  XDINS detected  in the optical  band, the second by the \vlt. 
For the  most reasonable 
distance range  to the  X-ray source, the  measured brightness  of the
candidate counterpart  (B$\sim27.4$) is barely  compatible with purely
thermal  emission   from  the  neutron  star  surface, while, assuming a 
value of $\dot E$ similar to that of other XDINSs,  
rotation-powered  emission from  the magnetosphere (eventually 
in combination with  a thermal component) requires \ns\ to have a 
very large 
optical efficiency ($\sim 3$ order of magnitudes larger than the Crab 
for $d=400$~pc). 
If the optical emission is powered by a
different process, the likely high magnetic field of
\ns\   tantalizingly  suggests  magnetar-like magnetospheric  emission 
as  a  viable option.   New  multi-band
observations, especially  in the  near-UV and in  the near-IR,  are 
required to characterize  the  counterpart spectrum  and  to assess  the
contribution of  possibly different  spectral components. At  the same
time, high resolution optical  astrometry of the candidate counterpart
with the  refurbished \hst\ could  yield the first direct  estimate of
the \ns\ parallax and distance,  crucial to build the neutron  star 
surface thermal
map,  and to provide confirmation of the optical counterpart 
through proper motion measurements.

\begin{acknowledgements}
SZ acknowledges support from a STFC (ex-PPARC) AF.
RM acknowledges STFC for support through a Rolling Grant. 
RT acknowledges financial support from ASI-INAF through grant AAE TH-058. 
LZ acknowledges financial support from ASI-INAF through
grant I/023/05/0.
\end{acknowledgements}

\clearpage 

%\clearpage 

\begin{thebibliography}{}

\bibitem[\protect\citeauthoryear{Bohlin et al.}{1978}]{bo78}
Bohlin, R.C., Savage, B. D., Drake, J. F. 1978, ApJ, 224, 132

\bibitem[\protect\citeauthoryear{Braje \& Romani}{2002}]{br02}
Braje, T.M., Romani, R.W. 2002, ApJ, 580, 1043

\bibitem[\protect\citeauthoryear{Caraveo et al.}{1998}]{car98}
Caraveo, P.A., Lattanzi, M.G., Massone, G., et al. 1998, A\&A, 329, L1

\bibitem[\protect\citeauthoryear{Cropper et al.}{2004}]{cr04}
Cropper, M. et al. 2004, MNRAS, 351, 1099

\bibitem[\protect\citeauthoryear{Cropper et al.}{2007}]{cr07}
Cropper, M. et al. 2007, Ap\&SS, 308, 161

\bibitem[\protect\citeauthoryear{Fitzpatrick}{1999}]{fitz99}
Fitzpatrick, E.L., 1999, PASP, 111, 63


\bibitem[\protect\citeauthoryear{Haberl}{2007}]{ha07}
Haberl, F. 2007, Ap\&SS, 308, 181

\bibitem[\protect\citeauthoryear{Heyl \& Kulkarni}{1998}]{hk98}
Heyl, J.S. \& Kulkarni, S.R. 1998, ApJ, 506, L61


\bibitem[\protect\citeauthoryear{Ho et al.}{2007}]{ho07}
Ho, W.C.G., et al. 2007, MNRAS, 375, 821

\bibitem[\protect\citeauthoryear{Israel et al.}{2003}]{isr03}
Israel, G.L. et al. 2003, ApJ, 589, L93

\bibitem[\protect\citeauthoryear{Kargaltsev  \&  Pavlov}{2007}]{kp07}
Kargaltsev, O.Y., Pavlov, G.G. 2007, Ap\&SS, 308, 287


\bibitem[\protect\citeauthoryear{Kaplan
et~al.}{2003a}]{kap03a}
Kaplan, D.L., Kulkarni, S., Van Kerkwijk, M.H. 2003a, ApJ, 588, L33

\bibitem[\protect\citeauthoryear{Kaplan
et~al.}{2003b}]{kap03b}
Kaplan, D.L., et al. 2003b, ApJ, 590, 1008


\bibitem[\protect\citeauthoryear{Kaplan
\& van Kerkwijk}{2005a}]{kap05a}
Kaplan, D.L., van Kerkwijk, M.H.  2005a, ApJ, 628, L45 


\bibitem[\protect\citeauthoryear{Kaplan
\& van Kerkwijk}{2005b}]{kap05b}
Kaplan, D.L., van Kerkwijk, M.H.  2005b, ApJ, 635, L65 

\bibitem[\protect\citeauthoryear{Kaplan}{2008}]{kap08}
Kaplan, D.L. 2008, proceedings of "40 Years of Pulsars: Millisecond Pulsars,
Magnetars, and More", August 12-17, 2007, McGill University,   
Montreal, Canada, AIP, 983, 331 

\bibitem[\protect\citeauthoryear{Kondratiev
et~al.}{2008}]{kon08}
Kondratiev, V.I. et al.  2008, submitted to ApJ 



\bibitem[\protect\citeauthoryear{Landolt}{1992}]{lan92}
Landolt, A.U. 1992, AJ, 103, 340 

\bibitem[\protect\citeauthoryear{Lasker et al.}{2008}]{las08}
Lasker, B.M. et al. 2008, submitted to AJ

\bibitem[\protect\citeauthoryear{Malofeev et al.}{2005}]{malo05}
Malofeev, V.M. et al. 2005, Astr. Rep., 49, 242

\bibitem[\protect\citeauthoryear{Malofeev et al.}{2006}]{malo06}
Malofeev, V.M. et al. 2006, ATel \#798;


\bibitem[\protect\citeauthoryear{McLaughlin et al.}{2006}]{mcl06}
McLaughlin, M.A. et al. 2006, Nature, 439, 817

\bibitem[\protect\citeauthoryear{McLaughlin et al.}{2007}]{mcl07}
McLaughlin, M.A. et al. 2007, ApJ, 670, 1307

\bibitem[\protect\citeauthoryear{Mereghetti}{2008}]{san08}
Mereghetti, S. 2008, arXiv:0804.0250, submitted to Astromomy and 
Astrophysics Review


\bibitem[\protect\citeauthoryear{Mignani et al.}{2004}]{mig04}
Mignani, R.P., De Luca, A., Caraveo, P.A. 2004, in Proc. of ``Young 
Neutron 
Stars and Their Environments'', IAU Symp. 218, eds. F. Camilo and B. 
Gaensler, ASP Conf. Proc., p. 391


\bibitem[\protect\citeauthoryear{Mignani et al.}{2007a}]{mig07a}
Mignani, R.P. et al. 2007a, AP\&SS, 308, 203


\bibitem[\protect\citeauthoryear{Mignani 
et~al.}{2007b}]{mig07b} 
Mignani, R.P., Zharikov,  S., Caraveo,  P.A. 2007b, A\&A, 473, 891

\bibitem[\protect\citeauthoryear{Mignani 
et~al.}{2007c}]{mig07c} 
Mignani, R.P. et al. 
2007c, A\&A, 471, 265



\bibitem[\protect\citeauthoryear{Motch et al.}{2003}]{mo03}
Motch, C., Zavlin, V.E., Haberl, F. 2003, A\&A, 408, 323

\bibitem[\protect\citeauthoryear{Motch et al.}{2005}]{mo05}
Motch, C., et al. 2005, A\&A, 429, 257


\bibitem[\protect\citeauthoryear{Pacini \& Salvati}{1983}]{ps83}
Pacini, F., Salvati, M. 1983, ApJ, 274, 369

\bibitem[\protect\citeauthoryear{Paresce}{1984}]{par84}
Paresce, F. 1984, AJ, 89, 1022


\bibitem[\protect\citeauthoryear{Pons et al.}{2002}]{pons02}
Pons, J.A., et al. 2002, ApJ, 564, 981

\bibitem[\protect\citeauthoryear{Popov et al.}{2006}]{pop06}
Popov, S.B., Turolla, R., Possenti, A. 2006, MNRAS, 369, L23 


\bibitem[\protect\citeauthoryear{Posselt et al.}{2007}]{pos07}
Posselt, B. et al. 2007, Ap\&SS, 308, 171


\bibitem[\protect\citeauthoryear{Predehl \& Schmitt}{1995}]{pr95}
Predehl, P. \& Schmitt, J.H.M.M. 1995, A\&A 293, 889

\bibitem[\protect\citeauthoryear{Rea et al.}{2007}]{rea07}
Rea, N. et al. 2007, MNRAS, 379, 1484

\bibitem[\protect\citeauthoryear{Severgnini et al.}{2005}]{seve057}
Severgnini, P. Della Ceca, R., Braito, V., et al., 2005, A\&A, 431, 87

\bibitem[\protect\citeauthoryear{Skrutskie et al.}{2006}]{Skrutskie06}
Skrutskie, M. F., et al. 2006, AJ, 131, 1163

\bibitem[\protect\citeauthoryear{Tr\"umper et al.}{2004}]{tru04}
Tr\"umper, J.E., Burwitz, V., Haberl, F., Zavlin, V.E. 2004, Nuclear 
Physics B Proceedings Supplements, 132, p. 560-565.

\bibitem[\protect\citeauthoryear{Van Kerkwijk \& Kaplan}{2007}]{vkk07}
van Kerkwijk, M.H., Kaplan, D.L. 2007, Ap\&SS, 308,
191

\bibitem[\protect\citeauthoryear{Van Kerkwijk \& Kaplan}{2008}]{vkk08}
van Kerkwijk, M.H., Kaplan, D.L. 2008, ApJ, 673, L163

\bibitem[\protect\citeauthoryear{Zacharias et al.}{2004}]{zac04}
Zacharias, N., et al. 2004, AJ, 127, 3043

\bibitem[\protect\citeauthoryear{Zampieri et al.}{2001}]{zamp01}
Zampieri L., et al. 2001, A\&A, 378, L5

\bibitem[\protect\citeauthoryear{Zane et al.}{2004}]{za04}
Zane, S., Turolla, R.,  Drake, J.J. 2004, AdSpR, 33, pp. 531-536.

\bibitem[\protect\citeauthoryear{Zane et al.}{2005}]{za05}
Zane, S., et al., 2005, ApJ, 627, 397

\bibitem[\protect\citeauthoryear{Zane et al.}{2006}]{zane06}
Zane, S., et al. 2006, A\&A, 570, 619



\end{thebibliography}
\end{document}